\documentclass[aps,12pt,preprint,groupedaddress,showpacs,reprintnumbers,amsmath,amssymb,floatfix]{revtex4}
\usepackage{ulem} 
\usepackage{graphicx}
\usepackage{dcolumn}
\usepackage{bm}
\usepackage{epsf}
\usepackage{epsfig}
\usepackage{amsmath,amsfonts,amsthm, amssymb,latexsym}
\usepackage{enumerate}
\usepackage{color}

\newcommand{\beq}{\begin{equation}}
\newcommand{\eeq}{\end{equation}}
\newcommand{\bcn}{\begin{center}}
\newcommand{\ecn}{\end{center}}

\newcommand{\lsim}{\lower0.5ex\hbox{$\; \buildrel < \over \sim \;$}}

\begin{document}

\title{Braking Index of Isolated Pulsars}

\date{\today}

\author{O.Hamil} \email{ohamil@vols.utk.edu}
\affiliation{Department of Physics and Astronomy, University of Tennessee, Knoxville, TN 37996, USA}
\author{J.R.Stone} \email{j.stone@physics.ox.ac.uk}
\affiliation{Department of Physics and Astronomy, University of Tennessee, Knoxville, TN 37996, USA}
\affiliation{Department of Physics, Oxford University, Oxford UK}
\author{Martin Urbanec} \email{martin.urbanec@fpf.slu.cz}\author{Gabriela Urbancov\'a} \email{gabriela.urbancova@fpf.slu.cz}
\affiliation{Institute of Physics, Faculty of Philosophy and Sciences, Silesian University in Opava, CZ 74601 Opava, Czech Republic}

\begin{abstract}
Isolated pulsars are rotating neutron stars with accurately measured angular velocities $\Omega$, and their time derivatives that show unambiguously that the pulsars are slowing down. Although the exact mechanism of the spin-down is a question of detailed debate, the commonly accepted view is that it arises through emission of magnetic dipole radiation (MDR) from a rotating magnetized body. Other processes, including the emission of gravitational radiation, and of relativistic particles (pulsar wind), are also being considered. The calculated energy loss by a rotating pulsar with a constant moment of inertia is assumed proportional to a model dependent power of $\Omega$. This relation leads to the power law $\dot{\Omega}$ = -K $\Omega^{\rm n}$ where $n$ is called the braking index. The MDR model predicts $n$ exactly equal to 3. Selected observations of isolated pulsars provide rather precise values of $n$, individually accurate to a few percent or better, in the range 1$ <$ n $ < $ 2.8, which is consistently less than the predictions of the MDR model. In spite of an extensive investigation of various modifications of the MDR model, no satisfactory explanation of observation has been found yet.

The aim of this work is to determine the deviation of the value of $n$ from the canonical $n = 3$ for a star with a frequency dependent moment of inertia in the region of frequencies from zero (static spherical star) to the Kepler velocity (onset of mass shedding by a rotating deformed star), in the macroscopic MDR model. For the first time, we use microscopic realistic equations of state (EoS) of the star to determine its behavior and structure. In addition, we examine the effects of the baryonic mass M$_{\rm B}$ of the star, and possible core superfluidity, on the value of the braking index within the MDR model.

Four microscopic equations of state are employed as input to two different computational codes that solve Einstein's equations numerically, either exactly or using the perturbative Hartle-Thorne method, to calculate the moment of inertia and other macroscopic properties of rotating neutron stars.  The calculations are performed for fixed values of M$_{\rm B}$ (as masses of isolated pulsars are not known) ranging from $1.0 - 2.2 M_\odot$, and fixed magnetic dipole moment and inclination angle between the rotational and magnetic field axes.  The results are used to solve for the value of the  braking index as a function of frequency, and find the effect of the choice of the EoS, M$_{\rm B}$. The density profile of a star with a given M$_{\rm B}$  is calculated to determine the transition between the crust and the core and used in estimation of the effect of core superfluidity on the braking index.

Our results show conclusively that, within the model used in this work, any significant deviation of the braking index away from the value $n=3$ occurs at frequencies higher than about ten times the frequency of the slow rotating isolated pulsars most accurately measured to date. The rate of change of $n$ with frequency is related to the softness of the EoS and the M$_{\rm B}$ of the star as this controls the degree of departure from sphericity. Change in the moment of inertia in the MDR model alone, even with the more realistic features considered here, cannot explain the observational data on the braking index and other mechanisms have to be sought.
\end{abstract}


\maketitle

\section{Introduction}\label{intro}
The slowing down of rotating neutron stars has been observed and modeled for decades. The simplest  models relate the loss the kinetic rotational energy of the star to the emission of magnetic radiation from a rotating dipolar magnetic field (MDR), attached to the star \cite{pacini1967, pacini1968, gold1968, gold1969,goldwire1969}. The calculated energy loss by a rotating pulsar is assumed proportional to a model dependent power of $\Omega$. This relation leads to the power law $\dot{\Omega}$ = -K $\Omega^{\rm n}$ where $n$ is called the braking index. The value of $n$ can, in principle, be determined from observation of higher-order frequency derivatives related to n by \cite{lyne2015}
\begin{eqnarray}
n& = & {{{\Omega} \ddot{\Omega}} \over {{\dot{\Omega}}^2}} \\
n(2n-1)& =& {{{\Omega}^2 \dddot{\Omega}} \over {{\dot{\Omega}}^3}}.
\end{eqnarray}

When the star is taken as a magnetized sphere, rotating in vacuum, with a constant moment of inerta (MoI) and a constant magnetic dipole moment, misaligned at a fixed angle to its axis of rotation, $n$ is equal to 3 (for derivation see Sec.~\ref{subsec:stat}). 

Extraction of the rotational frequency and its time derivatives from observation involves a detailed analysis of the time evolution of the pulses, and of the spectra and luminosity of radiation from the related nebulae in a wide range of wavelengths. Although data on many pulsars are available in the literature, there are only eight pulsars generally accepted to yield reliable data on the pulsar's spin-down (see Table.~\ref{tab1}, recent compilation \cite{magalhaes2012} and Refs. therein). The third derivative is known only for the Crab pulsar \cite{lyne1988}, and PSR B1509-58 \cite{kaspi1994}.

Examination of Table~\ref{tab1} shows that $n = 3$ does not agree with observation. There have been many attempts to extend/modify the basics of the MDR model.  These include consideration of magnetic field activity (e.g., \cite{livingstone2011,blanford1988, melatos1997, lyne2004,harding1999,kramer2006, lyne2010}), superfluidity and superconductivity of the matter within pulsars (e.g., \cite{sed1998, ho2012, page2014}), and modifications of the power law and related quantities (e.g., \cite{johnston1999, magalhaes2012}). Time dependence of the constants in the MDR model has also been considered \cite{blanford1988, contopoulos2006, zhang2012, gourgouliatos2014}. In particular, time evolution of the inclination angle between spin and magnetic dipole axes has been recently addressed \cite {lyne2015,lyne2013}. However, there is no model currently available that would yield, consistently, the typical spread of values of $n$ as illustrated in Table~\ref{tab1}.

Models outside the MDR have also been introduced. For example, energy loss through emission of accelerated charged particles, forming a massive wind from the surface of rotating stars \cite{michel1969, harding1999}, or emission of higher multipole electromagnetic radiation, including the gravitational quadrupole component \cite{ostriker1969,alford2014}, has been studied. Competition of different mechanisms, MDR, emission of gravitational waves, and particle winds was also investigated by \cite{carraminana1996, alvarez2004}.

In this work we focus on determination of the maximum deviation of the braking index from the value $n = 3$ by introducing two modifications of the simple MDR model: frequency dependence of MoI, related to the change of shape of a deformable star due to rotation, and superfluidity of the pulsar core. The correction to the expression for the braking index arising from these modifications following Glendenning \cite{glendenning} is derived, and included in the calculation, using four realistic equations of state (EoS) over a range of baryonic mass (M$_{\rm B}$) . We study the relation between the softness of the EoS and the rate of change of the braking index as a function of frequency and the M$_{\rm B}$. The four EoS were also used to obtain mass density profiles of the pulsars needed to determine the transition region between the crust and core. These results were utilized in the simulation of an effect of superfluid conditions that eliminates the angular momentum exchange at the threshold between the crust and core. The calculation is performed over a full range of frequencies of the pulsar from zero to the Kepler frequency and a range of M$_{\rm B}$ from 1.0 to 2.2 M$_\odot$, representing the gravitational mass range from about 0.8 to 2.0 M$_\odot$. 

The paper is organized as follows: The braking index, as calculated in the canonical MDR model, and its extensions introduced in this work, are presented Secs.~\ref{subsec:stat}~$-$~\ref{subsec:dyn}. Sec.~\ref{sec:calc} contains the computational method employed in this work, followed by Sec.~\ref{sec:res} where the main results are reported. The main findings of this work, discussion and outline of future development are summarized in Sec.~\ref{sec:concl}.

\subsection{Simple MDR model}
\label{subsec:stat}
The total energy loss by a rotating magnetized sphere can be expressed in terms of the time derivative of the radiated energy as \cite{pacini1968, glendenning, lyne2015}
\begin{eqnarray}
{dE \over dt} &=& -{2\over3}\mu^{\rm 2} {\Omega}^4 {\sin^2{\alpha}},
\label{eq:2}
\end{eqnarray}
where $\mu$ is the magnetic dipole moment of the pulsar, $\mu$ = B R$^3$. $R$ is the radial coordinate of a surface point with the surface magnetic field strength $B$, $\Omega$ is the rotational frequency, and $\alpha$ is the angle of inclination between the dipole moment and the axis of rotation \cite{glendenning}. 

Substituting the kinetic energy of a rotating body, dependent on the MoI $I$,
\begin{eqnarray}
E &=& {1\over2}I\Omega^2,
\label{eq:2.1}
\end{eqnarray}
into (\ref{eq:2}) yields
\begin{eqnarray}
{d\over dt}\left({1\over2}I\Omega^2\right) &=& -{2\over3}\mu^2 {\Omega}^4 {\sin^2{\alpha}}.
\label{eq:2.3}
\end{eqnarray}
Assuming constant MoI, $dI/dt$ = 0, we get
\begin{eqnarray}
\dot{\Omega} &=& -{2\over3}{{\mu^{\rm 2}}\over{I}} {\Omega}^3 {\sin^2{\alpha}}.
\label{eq:2.4}
\end{eqnarray}
Setting $K = {2\over3}{{\mu^2}\over{I}}  {\sin^2{\alpha}}$ in (\ref{eq:2.4}) and taking $\mu$ and $\alpha$ constant leads to the commonly used braking  power law describing the pulsar spin-down due to dipole radiation:
\begin{eqnarray}
\dot{\Omega} &=& -K\Omega^3.
\label{eq:2.5}
\end{eqnarray}
Differentiating (\ref{eq:2.5}) with respect to time
\begin{eqnarray}
\ddot{\Omega} &=& -3K\Omega^2\dot{\Omega},
\label{eq:2.6}
\end{eqnarray}
and combining (\ref{eq:2.5}) and (\ref{eq:2.6}) to eliminate $K$ we get the value of the braking index $n$
\begin{eqnarray}
n =  {{{\Omega} \ddot{\Omega}} \over {{\dot{\Omega}}^2}} &=& 3.
\label{eq:2.7}
\end{eqnarray}

\subsection{ MDR model with frequency dependent MoI}
\label{subsec:dyn}
The simple MDR value $n=3$ (\ref{eq:2.7}) is derived taking the  $I$, $\mu$, and $\alpha$ as independent of frequency and constant in time. However, in reality the MoI of rotating pulsars changes with frequency and, consequently with time. \cite{gle1997, glendenning}. The equlibrium state of a rotating pulsar includes the effect of centrifugal forces, acting against gravity. The shape of the pulsar is ellipsoidal with decrease (increase) in radius along the equatorial (polar) direction with respect to the rotation axis as the pulsar spins down. Thus, the MoI, and, consequently, the braking index, are both frequency dependent.

It is convenient to rewrite (\ref{eq:2.3}) as
\begin{eqnarray}
{d \over dt}\left({1\over2}I\Omega^2\right) &=& -C\Omega^4,
\label{eq:3}
\end{eqnarray}
\noindent
where $C= {2\over3}\mu^2 sin^2\alpha  $.
Assuming this time that $dI/dt$ is nonzero, differentiation of (\ref{eq:3}) with respect to time gives
\begin{eqnarray}
2I\dot{\Omega} + \Omega \dot{I} = -2C\Omega^3.
\label{eq:3.1}
\end{eqnarray}
Differentiating once more gives
\begin{eqnarray}
2I\ddot{\Omega} + 2\dot{\Omega}\dot{I} + \dot{\Omega}\dot{I} + \Omega\ddot{I} = - 6C\Omega^{2}\dot{\Omega}.
\label{eq:3.2}
\end{eqnarray}
Using the chain rule we can write $\dot{I}$ in terms of $\dot{\Omega}$
\begin{eqnarray}
{dI \over dt} = {d\Omega \over dt}{dI \over d\Omega},
\end{eqnarray}
and obtain
\begin{eqnarray}
\dot{I} = I'\dot{\Omega}\\ \ddot{I} = {\dot{\Omega}}^2I'' + I'\ddot{\Omega},
\end{eqnarray}
where the primed notation represents the derivative with respect to $\Omega$.

Substituting the identities shown above into (\ref{eq:3.1})~$-$~(\ref{eq:3.2}), we get the following relations for $\dot{\Omega}$ and $\ddot{\Omega}$,
\begin{eqnarray}
\dot{\Omega} &=& -{2C\Omega^2} \over {(2I + \Omega I')} \\
\ddot{\Omega} &=& {{-6C\dot{\Omega}\Omega^{2} -\dot{\Omega}^2(3I'+\Omega I'')} \over {(2I + \Omega I')}}.
\end{eqnarray}
After some algebra it is easy to show that the expression of the braking index as a function of angular velocity reads
\begin{eqnarray}
n(\Omega) &=&{{\Omega \ddot{\Omega}} \over \dot{\Omega}^2} = 3 -{{(3\Omega I' + \Omega^2 I'')} \over {(2I + \Omega I')}}.
\label{eq:3.3}
\end{eqnarray}

We note that the magnetic dipole moment of the nonspherical pulsar may, in principle, also change with frequency. Estimation of this effect would require knowledge of the origin and distribution of the dipole moment, which is lacking. We therefore ignore such a change here and restrict ourselves to analysis of the two effects described.

\section{Calculation Method} \label{sec:calc}
Previous modeling of the braking index using the simple MDR model with constant MoI assumed a pulsar with $1.4 M_{\odot}$ gravitational mass and a radius $\sim$ 10 km. In this work, which includes frequency dependent MoI and varying M$_{\rm B}$, we solve the equations of motion of rotating stars with realistic EoS using two different numerical methods.

\subsection{The codes}
The PRNS9 code, developed by Weber \cite{book:weber, weber_private}, is based on a perturbative approach to the equations of motion of  slowly rotating near-spherical objects \cite{hartle1968, hartle1973}. To ensure the reliability of the PRNS9 code results, we also used the RPN code. This code by Rodrigo Negreiros \cite{negreiros} is based on a publicly available algorithm, RNS, developed by Stergioulas and Friedman \cite{RNS}. The equations of motion are derived directly from Einstein's equations, following the  Cook, Shapiro and Teukolsky approach \cite{CST}, described in detail in \cite{komatsu1989}. Both codes are applicable to rotating stars with all frequencies up to the Kepler limit.

\begin{figure}
 \centering
 \vspace{1.0cm}
 \includegraphics[width=1.0\textwidth]{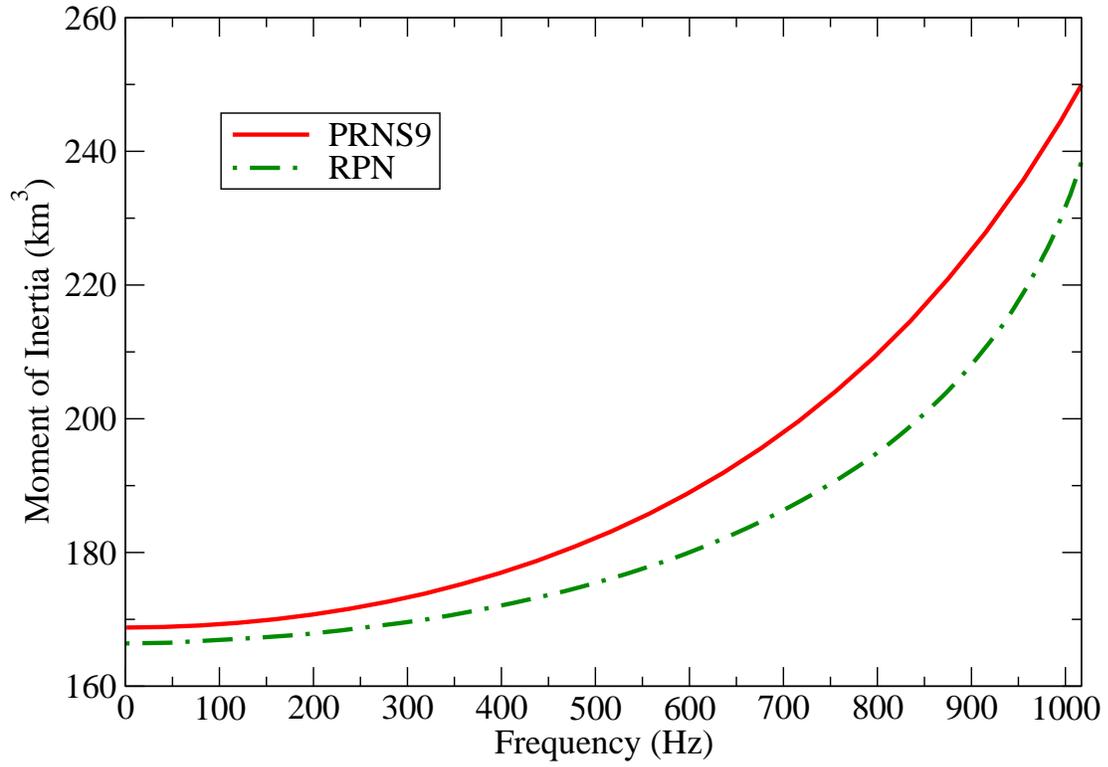}
 \caption{\label{fig1} MoI as a function of frequency for a pulsar with M$_{\rm B}$  = 2.0~$M_\odot$ as calculated with both RPN and PRNS9 numerical codes.}
 \end{figure}

A comparison of the results of the two codes is demonstrated in Fig.~\ref{fig1} which shows MoI as a function of frequency for a pulsar with the QMC700 EoS and M$_{\rm B}$ = 2.0 M$_\odot$ (see Sec.~\ref{subsec:eos}). They differ most, but by less than 10\%, as the Kepler frequency is approached. This difference has no consequence in practical applications as no pulsars rotating with frequencies in the Kepler region (see Table~\ref{tab2} for EoS and M$_{\rm B}$ used in this paper) have been observed. The small difference at near zero frequency (about 1.25\%), due to the difference in behavior of the two low density EoS (see Sec.~\ref{subsec:eos} ), is negligible in the context of calculating neutron star macroproperties.  

Unless stated otherwise, present the PRNS9 code results as this code provides calculation of some observables, not readily available in the RPN code, which are needed in our study (see Sec.~\ref{sec:res}).

\subsection{The equation of state} \label{subsec:eos}
An essential input to the calculation of macroscopic properties of rotating neutron stars is the EoS. The EoS is constructed for two physically different regimes, the high density core and the relatively low density crust.

\begin{figure}
 \centering
 \includegraphics[width=1.0\textwidth]{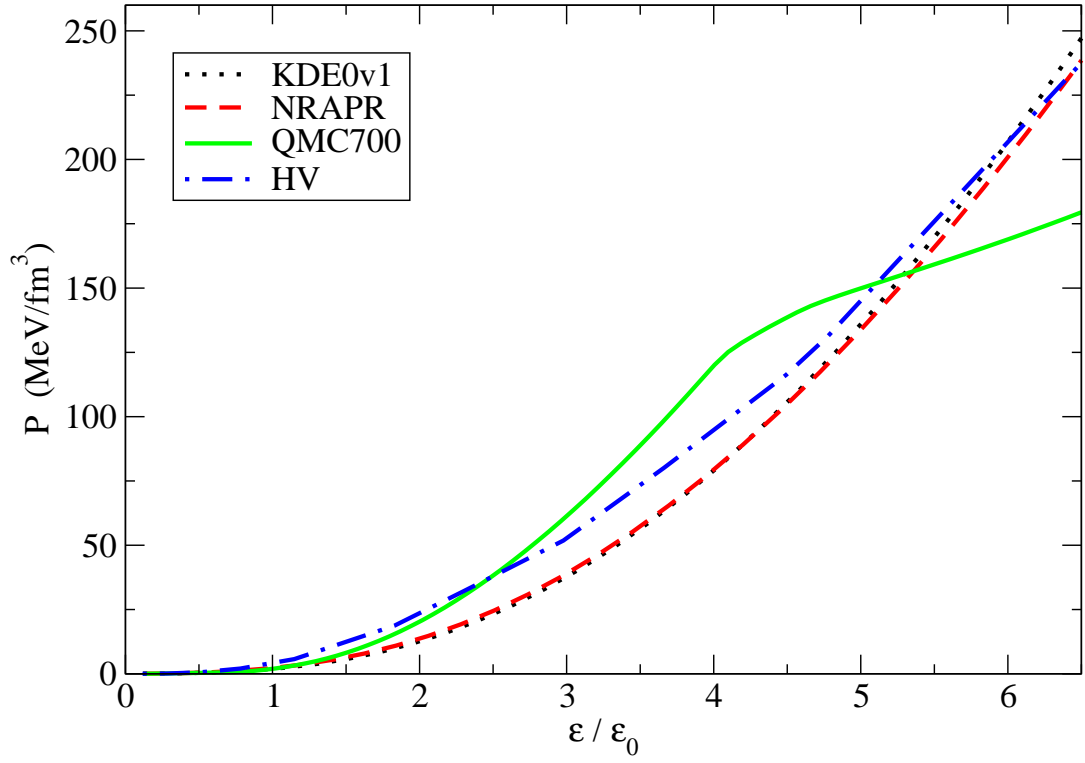}
 \caption{\label{fig2} Pressure vs energy density $\epsilon$ (in units of the energy density of symmetric nuclear matter at saturation $\epsilon_{\rm 0}$) as predicted by the four EoS used in this work.}
 \end{figure}

The microscopic composition of high density matter in the cores of neutron stars is not well understood. We have chosen two EoS, which assume that the core is made only of nucleons, KDE0v1 \cite{agrawal2005} and NRAPR \cite{steiner2005}. These EoS were selected by Dutra \textit{et al.} \cite{dutra2013} as being among the very few that satisfied an extensive set of experimental and observational constraints on properties of high density matter. In addition, we use two more realistic EoS that include in the core the heavy strange baryons (hyperons) as well as nucleons. The QMC700 EoS has been derived in the framework of the Quark-Meson-Coupling (QMC) model \cite{guichon2006, stone2007} and the Hartree V (HV) EoS \cite{weber1989} is based on a relativistic mean-field theory of nuclear forces.  The maximum mass of a static star, calculated using the Tolman-Oppenheimer-Volkof equation, is 1.96, 1.93, 1.98, and 1.98 M$_\odot$ for KDE0v1, NRAPR, QMC700, and HV, respectively, which is close to the gravitational mass of the  heaviest known neutron stars \cite{demorest2010, antoniadis2013}.  The EoS are illustrated in Fig.~\ref{fig2}, which shows pressure as a function of energy density $\epsilon$ in units of nuclear saturation energy density $\epsilon_0$ = 140 MeV/fm$^{\rm 3}$.  We observe that the pressure increases as a function of energy density almost monotonically for KDE0v1, NRAPR and HV, whereas QMC700 EoS predicts a change in the rate of increase at about 4 $\epsilon_0$. This change, and the subsequent softening of the EoS, happens at the transition energy density marking the threshold for appearance of hyperons in the matter. Such a change is not observed in the HV EoS. The main reason for the difference between the two hyperonic models is that the QMC700 distinguishes between the nucleon-nucleon and nucleon-hyperon interactions (neglecting the poorly known hyperon-hyperon interaction), whereas the HV model uses a universal set of parameters for all hadrons. Inclusion of both the QMC700 and HV EoS in this work reflects the uncertainty in the theory of dense matter in the cores of neutron stars.  

In order to ease the numerical calculation within the different architecture of the RPN and PRNS9 codes, two different EoS in the low density region (neutron star crust) have been used.  The Baym-Bethe-Pethick (see \cite{bbp} and Table V in \cite{bps}) at about 0.1 fm$^{\rm -3}$, which is augmented by Baym-Pethick-Sutherland \cite{bps} at about 0.0001 fm$^{\rm -3}$, going down to $\sim$ 6.0$\times$ 10$^{\rm -12}$  fm$^{\rm -3}$, was adopted in the RPN code.  For the PRNS9 code, the high density EoS was matched to the Harrison-Wheeler \cite{hw}, taken for the outer crust of the star, and Negele-Vautherin \cite{nv} EoS for the inner crust.

\section{Results and discussion}\label{sec:res}
As detailed in the previous section, the calculation of the frequency dependence of the braking index has been done for a multiple combination of codes, EoS's and M$_{\rm B}$ of the rotating star. We show only typical examples of the results, usually for the QMC700 EoS, unless stated otherwise.

\subsection{Braking index with frequency dependent MoI}
\begin{figure}
 \centering
 \includegraphics[width=1.0\textwidth]{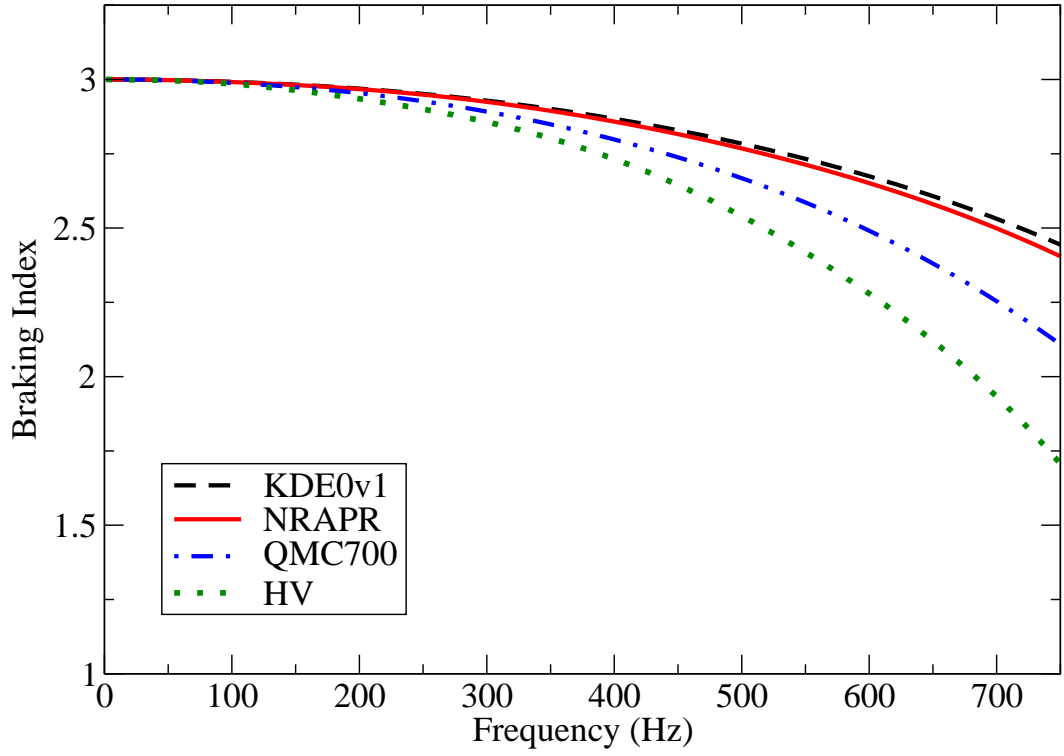}
 \caption{\label{fig3} Braking index as a function of frequency calculated for a pulsar with M$_{\rm B}$  = 2.0~$M_\odot$ with all EoS adopted in this work.}
 \end{figure}

\begin{figure}
 \centering
 \vspace{1.0cm}
 \includegraphics[width=1.0\textwidth]{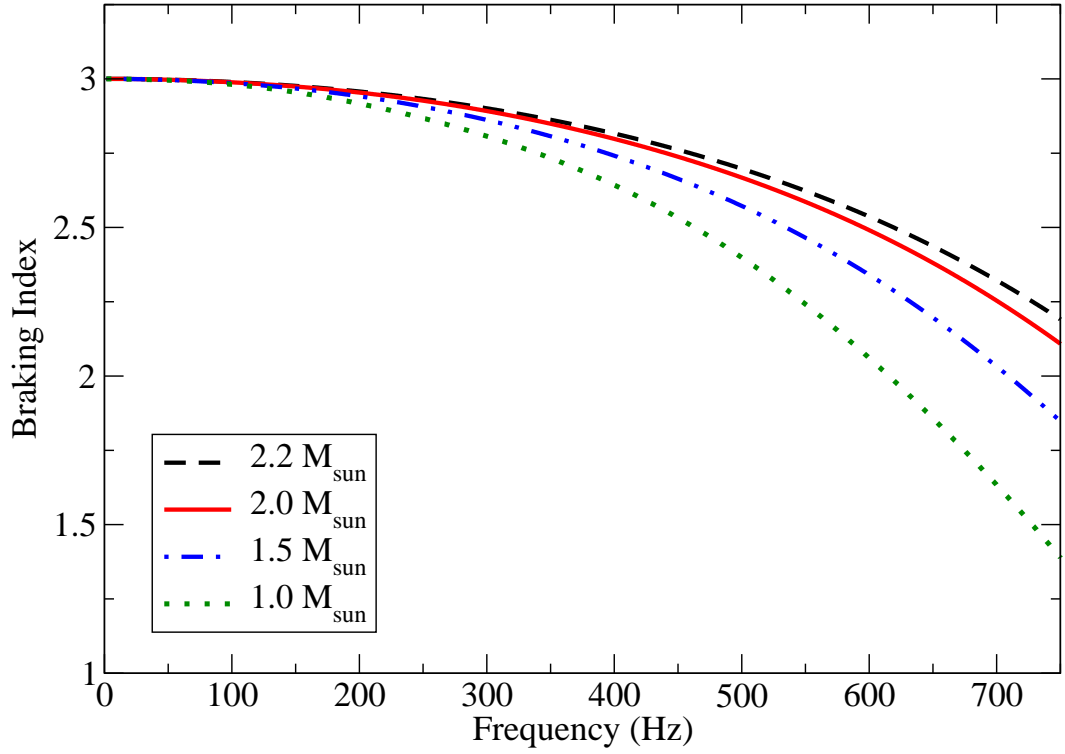}
 \caption{\label{fig4} Braking index as a function of frequency calculated of pulsars with M$_{\rm B}$ =  1.0 -  2.2 M$_\odot$.}
 \end{figure}

As a general feature, we find that any appreciable deviation of the braking index from the generic value $n = 3$ is observed only at rotational frequencies higher than about 250 Hz. The sensitivity of this deviation to the EoS and M$_{\rm B}$ is demonstrated in Figs.~\ref{fig3}~$-$~\ref{fig4}. As can be seen in  Fig.~\ref{fig3},  the biggest change in the braking index of a 2.0 M$_\odot$ star pulsar is predicted by the HV EoS, followed by the QMC700, reaching $\sim$ values 1.75 and 2.15 at 750 Hz, respectively. The two nucleon-only EoS, KDE0v1 and NRAPR, behave in a very similar way and predict a larger value of n = 2.5 at this frequency. These trends can be directly related to the properties of the EoS's. Fig.~\ref{fig4} shows the sensitivity to  M$_{\rm B}$ for the QMC700 EoS. The effect clearly increases with decreasing  M$_{\rm B}$.

\begin{figure}
 \centering
 \vspace{1.0cm}
 \includegraphics[width=1.0\textwidth]{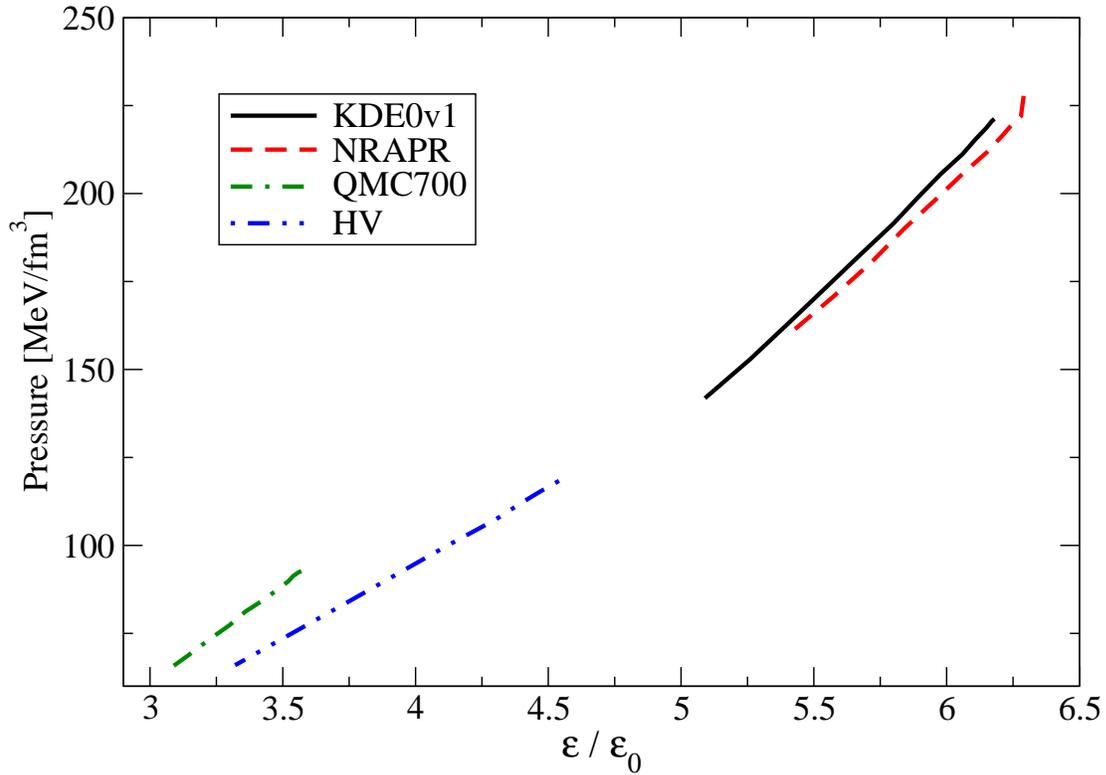}
 \caption{\label{fig5} Pressure as a function central density of a pulsar with M$_{\rm B}$  = 2.0  M$_\odot$, rotating with frequencies decreasing from the Kepler limit to zero as predicted by the four EoS used in this work. Relation between the central density and frequency is given in  Table~\ref{tab3}. For more explanation see text. }
 \end{figure}

We recall that the stiffness/softness of the EoS relates to the rate of change of pressure with changing energy density $\epsilon$. For each EoS the frequency dependent $\epsilon$ is given in Table~\ref{tab3}. Using Fig.~\ref{fig2} we obtain the corresponding pressures. Figure~\ref{fig5} plots the pressure vs $\epsilon$ relation for each EoS, the slope indicating the stiffness/softness in each case. The maximum change in the braking index in Fig.~\ref{fig3}, observed for the HV EoS,
is seen to be associated with the smallest change in pressure with increasing $\epsilon$, i.e., the largest softness. The stiffest EoS's,  KDE0v1 and NRAPR, predict the smallest response to the pulsar's rotational deformation. This conclusion is further supported by results shown in Fig.~\ref{fig4}. Pulsars with the lowest M$_{\rm B}$, governed by the softest EoS, exhibit the largest change in the braking index at high frequency. 

\subsection{Superfluidity of the core}
The effects demonstrated in Figs.~\ref{fig3}~$-$~\ref{fig4} were calculated assuming that the whole body of a pulsar contributes to the total (core+crust) MoI. However, some theories suggest the conditions inside a pulsar are consistent with the presence of superfluid/superconducting matter, both in the crust and in the core \cite{sed1998, ho2012, page2014, hooker2013}. Superfluid material would contribute to  the rotation, thus reducing the MoI. 

\begin{figure}
 \centering
 \vspace{1.0cm}
 \includegraphics[width=1.0\textwidth]{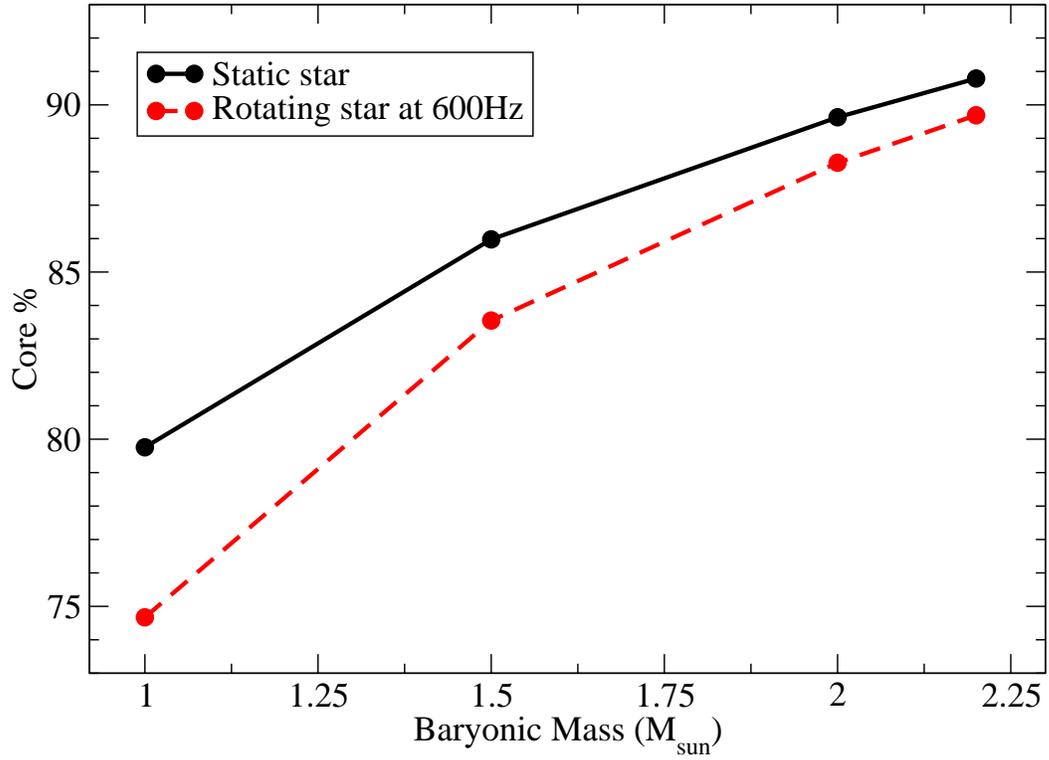}
 \caption{\label{fig6} Percentage of the core in a static star and a star rotating at 600 Hz as a function of M$_{\rm B}$. For more explanation see text. }
 \end{figure}

In this work we considered an extreme case in which the whole contribution of the core to the total MoI is removed. This scenario could be realized, for example, if either the whole core is superfluid or there is a layer of superfluid material between the core and the inner crust of the star, preventing an angular momentum transfer between the core and the crust. Either scenario simply results in removal of the contribution of the core to the MoI. To model this effect, it was necessary to locate the transition between the two phases of neutron star matter, which is assumed to occur roughly at 120 MeV/fm$^{\rm 3}$. The corresponding pressure is dependent on the EoS and is used to locate the physical position of the transition along the equatorial radius. The results are schematically illustrated in Fig.~\ref{fig6}, which shows the proportion of the core radius in a static star and a star rotating at an arbitrarily chosen frequency of 600 Hz as a function of M$_{\rm B}$, as predicted by the KDE0v1 EoS.  We observe that the proportion is different for a static and rotating star, mainly because of the varying shape of the star. The expansion (compression) along the equatorial (polar) direction with respect to the rotational axis leads to larger deformation of the less dense crust than that of the denser core.

\begin{figure}
 \centering
 \includegraphics[width=1.0\textwidth]{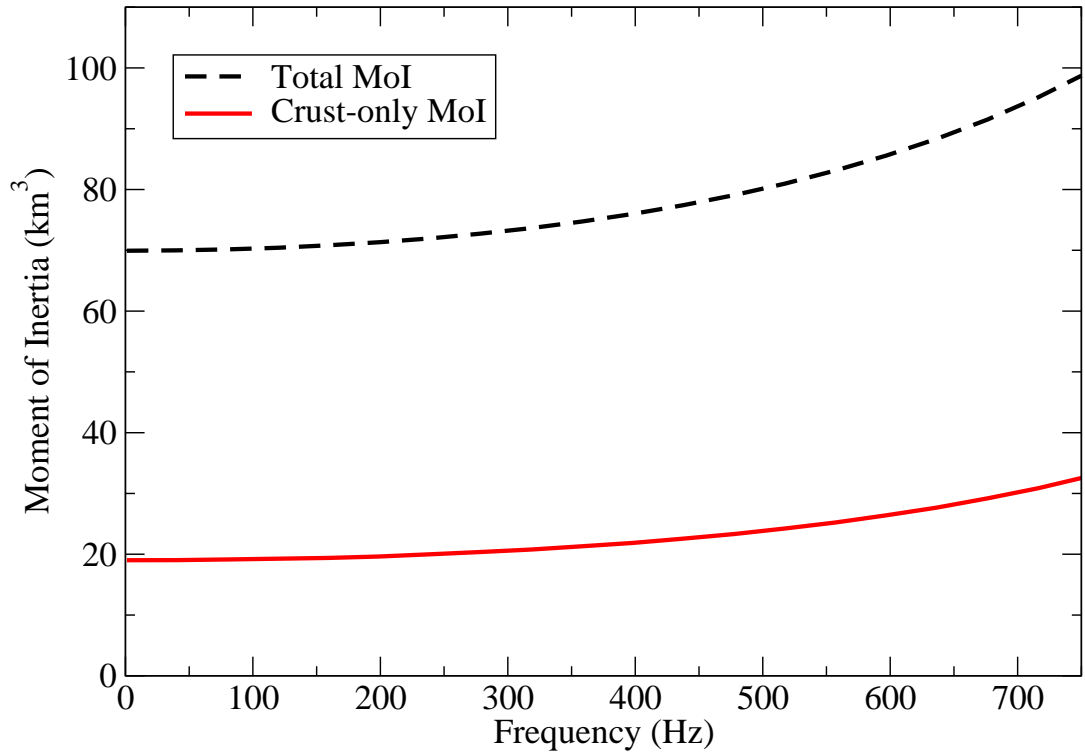}
 \caption{\label{fig7} Total (crust and core) and the crust-only MoI as a function of frequency, calculated for a pulsar with M$_{\rm B}$  = 1.0 M$_{\odot}$.}
 \end{figure}

\begin{figure}
\centering
\includegraphics[width=1.0\textwidth]{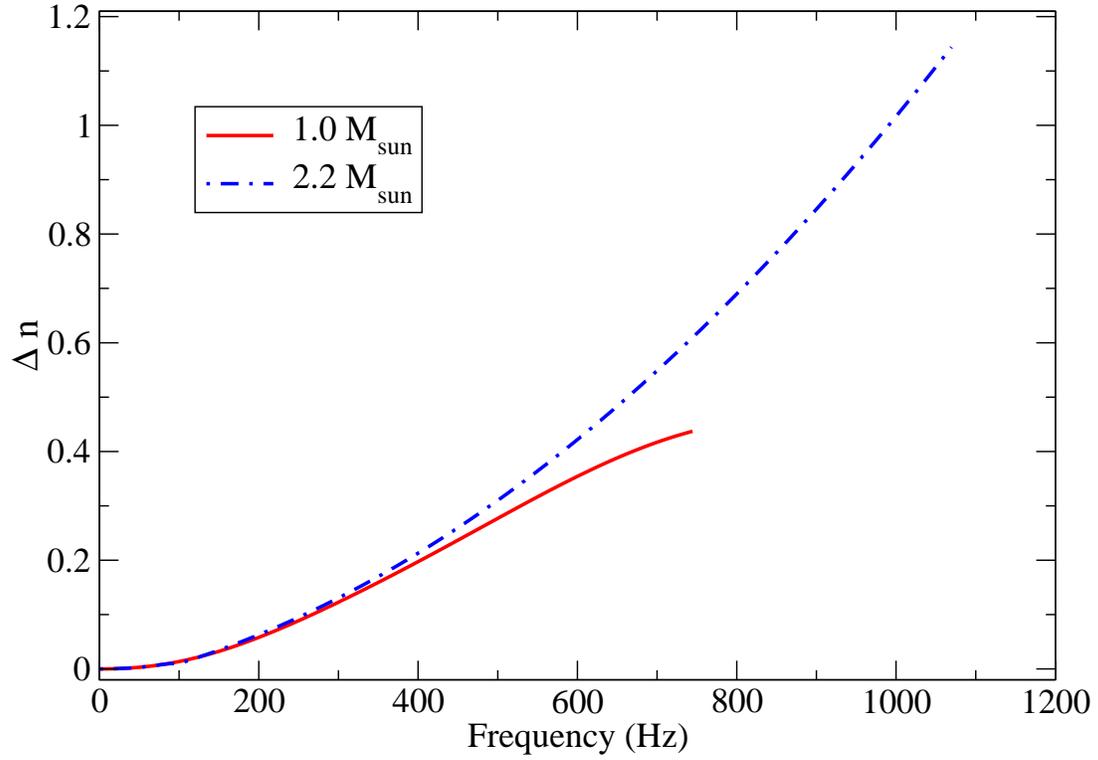}
\caption{\label{fig8} $\Delta$ n represent the difference in braking index as a function of frequency between stars with (see Fig.~\ref{fig4}) and without core contribution to the MoI.  Each curve is displayed up to the Kepler frequency of the star.}
\end{figure}

Elimination of the core contribution can lead to a dramatic lowering of the  total MoI by more than a factor of three, as shown, for example, in Fig.~\ref{fig7} for the 1.0~M$_\odot$ baryon mass and the QMC700 EoS. The difference between the total and crust-only MoI shows a weak frequency dependence with a slight increase above about 600 Hz. In turn, the reduction of the MoI by removal of the core contribution leads to additional changes in the braking index, on top of the changes due to the frequency dependent MoI (see Fig.~\ref{fig4}) as shown in  Fig.~\ref{fig8}. This change is, as expected, larger for higher mass stars that contain a more significant proportion of dense core material than for lower mass stars that are more crustlike throughout.

For the purpose of this work, we assumed that there is substantial superfluid material located in the core of the star but made no assumption on the superfluidity (or not) of the crust. The effects reported here should be taken as illustrative rather than definitive of the possible effects of superfluidity.

\subsection{Summary} 
The variation of  braking index of isolated rotating neutron stars with M$_{\rm B}$ = 1.0, 1.5, 2.0, and 2.2~M$_\odot$  with rotational frequency from zero to the Kepler limit within the MDR model with frequency dependent MoI has been investigated. The microphysics of the star was included through utilizing realistic EoS of the pulsar matter. An illustration of the possible effect of superfluidity in the star core has been included in the study. 

\begin{figure}
\includegraphics[width=1.0\textwidth]{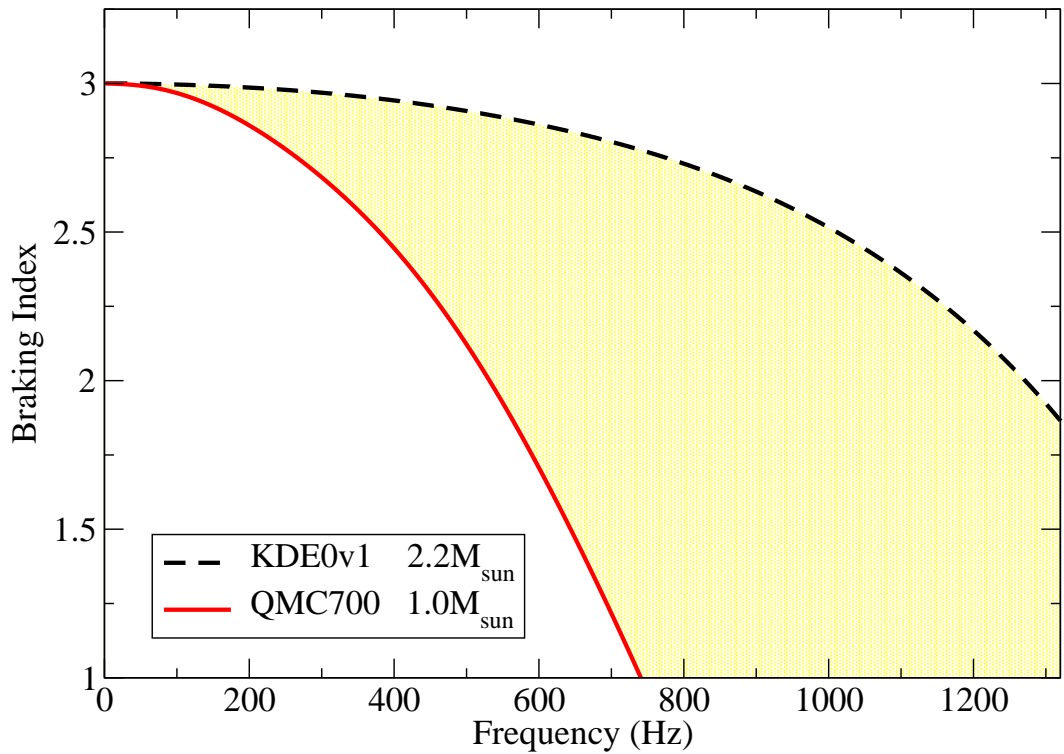}
\caption{\label{fig9} Lower and upper limits on values of the braking index as a function of frequency, including results from both numerical codes, all EoS, M$_{\rm B}$, and the superfluid condition. The (yellow) shaded area between the two lines defines the location of all results within the limits. The pulsar with baryonic M$_{\rm B}$ = 2.2  M$_{\odot}$ and the KDE0v1 EoS has the highest Kepler frequency (see Table ~\ref{tab2}) and defines the frequency limit in this work.}
\end{figure}

Compiling results of all models used in this work, including the superfluidity effect, we deduce a definitive upper and lower limit on the braking index as a function of frequency, shown in Fig.~\ref{fig9}.  The maximum change in the braking index is obtained with the QMC700 EoS and 1.0 M$_\odot$; the least effect is found for KDE0v1 and the 2.2 M$_\odot$ star.  Reduction of the braking index from the simple MDR model value $n = 3$ happens only at frequencies that are some significant fraction of the Kepler frequency. The calculation predicts that isolated pulsars with the braking index most deviating from $n = 3$ have low  M$_{\rm B}$. For the frequencies of known isolated pulsars with accurately measured braking indices (see Table~\ref{tab1}), the reduction away from $n=3$ found in this model, is negligible.

\section{Conclusions and outlook}\label{sec:concl}
\begin{figure}
\centering
\vspace{1.0cm}
\includegraphics[width=1.0\textwidth]{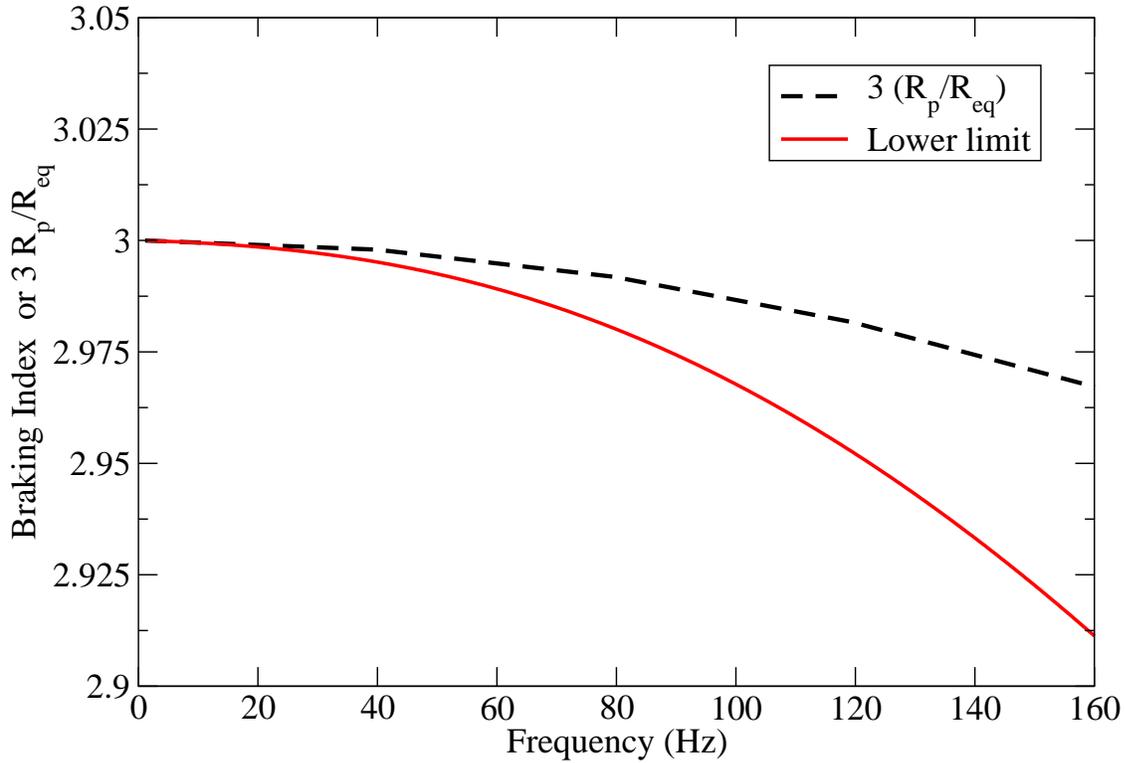}
\caption{\label{fig10} The lower limit of the braking index (see Figure~\ref{fig9}) as a function of frequency (solid line) compared with the ratio between polar (R$_{\rm p}$) and equatorial radii (R$_{\rm eq}$), normalized to three, which determines deformation of the star. The difference between the two lines represents a corelation between deviations of the braking index from  $n = 3$ and deformation for a 1.0 M$_{\odot}$ pulsar rotating at frequencies below 160 Hz (notice the expanded y-scale). It is seen that the shape deformation, even for this most deformable star, is small at these frequencies and quite unable to reproduce the observed range of braking indices.}
\end{figure}

In the model of isolated pulsars used in this work, the rate of change in rotational frequency of a spherical magnetized pulsar \textit{in vacuo} depends on three factors: the MoI (constant or frequency dependent), the magnitude of the magnetic dipole moment, and the inclination angle between the magnetic and rotational axes. The braking index is related to changes of these observables. In this work we considered effects due to changes to the MoI and its variation and showed that any significant deviation from the $n = 3$ value  appears only at frequencies much higher than the frequency range of observed isolated pulsars with reliable braking index. As MoI is related to the shape of the star, this result is consistent with the assumption of the simple MDR model that the pulsar is spherical at low frequencies. We show in  Fig.~\ref{fig10} the correlation between the braking index and the deformation of the star in terms of the polar to equatorial radii ratio R$_{\rm p}$/R$_{\rm eq}$ (normalized to 3 for display). It follows that changes in MoI alone cannot explain the observed deviation of the braking index at low frequencies  from the simple MDR model predictions. If the MDR model is to be  sustained, attention has to be paid to changes in the magnitude magnetic dipole moment and/or in the inclination angle. As stated above, the lack of knowledge of the origin and properties of the pulsar's magnetic field makes the former task difficult but the latter may be worth pursuing, particularly in the view of recent work by Lyne~\textit{et al.} \cite{lyne2013, lyne2015}.

Finally, we have shown that the simple exclusion of the core due to the superfluidity, or some superfluid  barrier between the crust and core, does not have a strong effect on braking in the frequency range of observed isolated pulsars. Further development of the idea of a macroscopic description of superfluidity would be interesting. Change of the magnetic field due to superfluidity and possible magnetic field expulsion, and a consequential increase in surface magnetic field strength $B$ could also be usefully explored.  

\bigskip

{\bf ACKNOWLEDGMENTS}
  
J.R.S. wishes to thank Jocelyn Bell Burnell for suggesting the topic of this work and helpful discussions, Fridolin Weber for help with understanding the PRNS9 code, and Nick Stone for discussions and advice concerning all of the physics relevant to this work. Discussions with John Miller are also gratefully acknowledged. O.H. is grateful for discussions with Rodrigo Negreiros, and use of the RPN code.  J.R.S. and O.H. thank the Institute of Physics in Opava for hospitality at the final stages of this work. G.U. and M.U. thank the Oak Ridge National Laboratory for kind hospitality. M.U. acknowledges support of the Czech Grant No. GA\v{C}R 209/12/P740 and Grant No.  CZ.1.07/2.3.00/20.0071 ''Synergy,'' aimed to foster international collaboration of the Institute of Physics of the Silesian University, Opava.  The research was also supported in part by the Department of Physics and Astronomy, University of Tennessee.

\clearpage
\centering
\begin{table}
\caption{\label{tab1} Selected pulsars adopted from \cite{magalhaes2012,espinoza2011,lyne2015}. $n_{\rm freq}$ is the value of $n$ obtained for the given pulsar, at the given frequency, with changing MoI calculated in this work.}
 \vspace{5pt}
\begin{tabular}{lcccl}
\hline  
PSR                           &       Frequency          &  $n$                            & $n_{\rm freq}$     & Ref.  \\
                                     &            (Hz)               &                                 &                         &     \\ \hline
 B1509$-$58              &     6.633598804    &    2.839$\pm$0.001    & 2.999       &  \cite{livingstone2007}  \\
 J1119$-$6127          &     2.4512027814   &   2.684$\pm$0.002    & 2.999       & \cite{waltevrede2011}   \\
 J1846$-$0258          &     3.062118502     &   2.65$\pm$0.1          & 2.999       &\cite{livingstone2007}     \\
                                &                             &   2.16$\pm$0.13         &                    &\cite{livingstone2011} \\
 B0531+21 (Crab)      &    30.22543701      &   2.51$\pm$0.01         &  2.995       &  \cite{lyne1993} \\
 B0540$-$69             &    19.8344965        &   2.140$\pm$0.009     & 2.997       & \cite{livingstone2007,boyd1995}  \\
 J1833$-$1034         &    16.15935711      &   1.8569$\pm$0.001    & 2.999   & \cite{roy2012}   \\
 B0833$-$45 (Vela)    &    11.2                   &   1.4$\pm$0.2            & 2.999        & \cite{lyne1996}   \\
 J1734$-$3333          &      0.855182765    &    0.9$\pm$0.2           & 3.000       &\cite{espinoza2011}   \\  \hline
\end{tabular}
\end{table}

\clearpage
\centering
\begin{table}
\caption{\label{tab2} Kepler frequencies in Hz for EoS and  M$_{\rm B}$/M$_\odot$ used in this work.}
 \vspace{5pt}
\begin{tabular}{lcccc}
\hline
EoS/M$_{\rm B}$             &   1.0    &   1.5     &     2.0   &2.2    \\ \hline
KDE0v1                       &  820    &   1024   &   1230  &   1327   \\                                       
NRAPR                        &   778   &    985    &   1198  &   1312    \\
QMC700                     &   745   &    888    &   1017  &   1070    \\
HV                             &   612   &    761    &   908    &   985      \\  \hline 
\end{tabular}
\end{table}

\clearpage
\centering
\begin{table}
\caption{\label{tab3} Central energy density (in units of the $\epsilon_{\rm 0}$ ) of a pulsar with HV,  QMC700,  NRAPR,  KDE0v1 EoS and M$_{\rm B}$  = 2.0 M$_\odot$ as a function of decreasing rotational frequency. For more explanation see text.}
 \vspace{5pt}
\begin{tabular}{lcccc}
\hline
Frequency [Hz] &  HV   &   QMC700  &   NRAPR   &  KDE0v1  \\  \hline
1200                  &   $-$    &       $-$     &      5.04       &         5.09  \\ 
1100                  &   $-$    &       $-$     &      5.23      &         5.26  \\ 
1000                  &   $-$    &       3.05   &       5.40      &        5.41  \\ 
900                    &    3.32  &       3.14   &       5.57      &        5.55  \\ 
800                    &    3.54  &       3.22   &       5.72      &         5.68  \\ 
700                   &     3.75  &       3.30   &       5.84     &          5.80 \\ 
600                   &     3.94  &   3.36   &      5.96     &         5.90 \\ 
500                  &    4.11  &    3.43    &     6.06     &      5.98 \\
400                   &   4.30  &    3.48    &     6.15     &    6.06   \\
 300                   &  4.40 &     3.52     &      6.21    &    6.11 \\
200                    &  4.47  &    3.54      &   6.25        &      6.15   \\
100                     &   4.52   &    3.56    &        6.28    &            6.17  \\
 0                        & 4.54 &    3.57       &    6.29   &   6.18      \\   \hline
\end{tabular}
\end{table}

\clearpage
\vskip 2cm

\clearpage
\vskip 2cm

\clearpage
\vskip 2cm

 \clearpage
\vskip 2cm
 
  \clearpage
\vskip 2cm
 
  \clearpage
\vskip 2cm
  
\clearpage
\vskip 2cm

\clearpage
\vskip 2cm

\clearpage
\vskip 2cm

\clearpage
\vskip 2cm

\end{document}